\documentstyle{article}
\begin{document}
\title{Elastic fields of stationary and moving dislocations
in finite samples}
\author{{\sc Rodrigo Arias} \\ 
Departamento de F\'\i sica, Facultad de Ciencias F\'\i sicas y  
Matem\'aticas \\ Universidad de Chile, Casilla 487-3, Santiago, Chile}
\maketitle
\newcommand{\beq}{\begin{equation}}
\newcommand{\eeq}{\end{equation}}
\newcommand{\bea}{\begin{eqnarray}}
\newcommand{\eea}{\end{eqnarray}}

\begin{abstract}
Integral expressions are determined for the elastic displacement and
stress fields due to stationary or moving dislocation loops in finite
samples. These general expressions are valid for anisotropic media as well. 
Specifically for the stress fields, a line integral representation is found, 
thus showing rigorously the independence of the stress fields
with respect to the choice of slip planes. In the stationary case the line
integral representation involves calculating 
a "vector potential" dependent on the specific geometry of the sample. 
Two examples of geometries, isotropic half space 
and thin plate, are shown where the "vector potential"
has been explicitly determined. With this general method one recovers some
earlier specific results in these geometries. 
\end{abstract}

\section{Introduction}
A general formula for
the displacement and stress fields generated by a dislocation loop
undergoing arbitrary motion in an infinite medium
was obtained some years ago by
Mura (1963). It is written
as a convolution of
the medium's impulse
response with a source localized along the dislocation loop
(i.e. independent of the loop's slip plane).
These formulae do not apply to the case of finite samples since
they are derived assuming homogeneity in space, i.e. a dependence
of the Green's function in $\vec{x}-\vec{x}'$, which indeed it is
no longer true in a finite sample.
The purpose of this paper is to 
announce a generalization of these formulae to
finite samples, as well as
explicit forms for the geometries of a half space and 
thin plate. The main results will be presented,
with some examples, but the detailed proofs will be
presented elsewhere.

Motivation for this work comes partly from an ongoing project that
attempts to understand recently observed dynamic instabilities for
cracks in thin plates (Sharon, Gross and Fineberg 1996, Boudet,
Ciliberto and Steinberg 1996) in terms of the interaction of the 
plate's oscillations with the crack's tip regarded as a continuous
distribution of infinitesimally small dislocations (Lund 1996). 
An intuitive thought behind this work was that the stress fields produced by
stationary or moving dislocation loops in a finite sample
should not depend on the slip planes chosen, and that indeed
the stresses will be continuous through the slip planes,
as it was in an infinite medium. This
is rigorously shown here: the expressions
found for the stress fields involve only line integrals over
the dislocation loops.

When the loops are stationary in order to find the explicit
form of the line integrals appearing in the stress fields, one
should calculate a "vector potential" appropriate to the finite
sample in question. This was done for the two examples shown
here: a half space and a thin plate. In a non stationary
situation the displacement velocity field can be written in
terms of line integrals, 
and from this one deduces that the
temporal partial derivative of the stress field can be written
in terms of them too. One finally deduces that the
time dependent stress field can be written in terms of line
integrals since one knows that the initial condition 
(stationary case) and its temporal derivative can both be 
written in terms of line integrals. 

\section{Displacement fields due to dislocations in finite samples}

\subsection{Stationary case}

The following formula is valid in a finite sample for the 
displacement field produced by a stationary dislocation loop
(derived in detail elsewhere).
It is written as an integration over the slip plane $S'$, and
involves a discontinuity of the displacement there, given by
the Burgers vector $b_{i}$:
\beq
U_{m}(x) = -b_{i}  \int_{S'}
dS_{j}' \sigma_{ij}^{Gm}(x',x) \: .
\label{ums}
\eeq
$\sigma_{ij}^{Gm}$ is the elastic stress associated with
the static Green's function $G_{lm}(x,x')$, which   
is the displacement (direction ($l$))
produced by a  
localized impulse force in space ($x'$) (direction ($m$)):
\beq
C_{ijkl}\frac{\partial^{2}}{\partial x_{k}
\partial x_{j}} G_{lm}(x,x') \equiv 
\frac{\partial}{\partial x_{j}} \sigma_{ij}^{Gm}(x,x') =
-\delta_{im}\delta(x-x') \: .
\label{eqs}
\eeq
It satisfies the free surface boundary condition:
\beq
\sigma_{ij}^{Gm}(x_{S},x')n_{j}(x_{S})=0 \: ,
\eeq
with $n_{j}(x_{S})$ the components of the normal to the
free surface at its point $x_{S}$.

\subsection{Non stationary case}

The following formula is valid in a finite sample for the
displacement field produced by a moving dislocation loop. 
It involves integration over past slip planes $S(t')$:
\beq
U_{m}(x,t) = -b_{i} \int_{-\infty}^{\infty} dt' \int_{S(t')}
dS_{j}' \sigma_{ij}^{Gm}(x',x;t-t') \: .
\eeq
The Green's function $G_{im}(x,x';t-t')$ 
is the displacement (in direction ($i$)) produced by a localized
impulse force (in direction ($m$)) in space ($x'$) and time ($t'$):
\beq
\rho \frac{\partial^{2} G_{im}}{\partial t^{2}} (x,x';t-t')
-\frac{\partial}{\partial x_{j}} \sigma_{ij}^{Gm}(x,x';t-t') =
\delta_{im}\delta(x-x')\delta(t-t') \: ,
\eeq
and satisfies that the normal stresses it produces at the free surface
of the sample are zero.

\section{Stress fields due to dislocation loops in finite samples}

\subsection{Stationary case}
Starting from Eq. (\ref{ums}), the
stress field produced by the static dislocation loop is:
\beq
\sigma_{pq}(x) = -b_{i}C_{pqlm} \int_{S'} dS_{j}' 
\frac{\partial}{\partial x_{l}} 
 \sigma_{ij}^{Gm}(x',x) \: .
\eeq
This can be expressed as a line integral along the dislocation loop:
\beq
\sigma_{pq}(x) = b_{i} C_{pqlm} \int_{L'} dl_{s}'
\{ \epsilon_{sjl} \sigma_{ij}^{Gm}(x',x) -
A_{s}^{ilm}(x',x) \} \: ,
\label{fin}
\eeq
with the "vector potential" $A_{s}^{ilm}(x',x)$ to be determined for
each specific sample. It is defined by:
\beq
\frac{\partial}{\partial x_{l}}
\sigma_{ij}^{Gm}(x',x)+\frac{\partial}{\partial x_{l}'}
\sigma_{ij}^{Gm}(x',x) =
\epsilon_{jrs} \frac{\partial}{\partial
x_{r}'} A_{s}^{ilm}(x',x) \: .
\label{curl}
\eeq

\subsection{Non stationary case}

Since the problem is
homogeneous under time translations it is possible to obtain an
expression for particle velocity, 
that involves an integral along the dislocation loop
only.
Indeed, 
from Eq. (\ref{ums}):
\beq
\frac{\partial U_{m}}{\partial t}(x,t)  = 
b_{i} \int_{-\infty}^{\infty} dt' \{ \frac{\partial}{\partial t'} [
\int_{S(t')} dS_{j}'  \sigma_{ij}^{Gm}
(x',x;t-t') ] 
 - \int_{\frac{dS}{dt'}(t')}
dS_{j}' \sigma_{ij}^{Gm}(x',x;t-t') \}
\: .
\eeq
The first term is zero since the Green's function vanishes at
$t-t'=\pm \infty$.
Since $\int_{dS(t')/dt'}dS_{j}' = \epsilon_{jpq} \int_{L(t')} dl_{q}'
V_{p}(x',t')$ ($L(t')$ is the dislocation loop bounding the slip
plane $S(t')$, $V_p(x',t')$ is the loop's local
 velocity and $\epsilon_{jpq}$ the completely
antisymmetric tensor in three dimensions), we have:
\begin{equation}
\frac{\partial U_{m}}{\partial t}(x,t) = -b_{i} 
\int_{-\infty}^{\infty} dt'
\int_{L(t')} dl_{q}' \sigma_{ij}^{Gm}(x',x;t-t') \epsilon_{jpq}
V_{p}(x',t') \: .
\end{equation}
From this expression for the displacement velocity field, one 
obtains an expression for the partial time derivative of the 
stress field that involves line integrals over the dislocation loop:
\begin{equation}
\label{dudt}
\frac{\partial}{\partial t}
\sigma_{pq}(x,t) = -C_{pqkm}b_{i} \int dt'
\int_{L(t')} dl_{q}' \frac{\partial}{\partial x_{k}} 
\sigma_{ij}^{Gm}(x',x;t-t') \epsilon_{jpq}
V_{p}(x',t') \: .
\end{equation}

As mentioned in the Introduction, this result together with the
result for a static loop (considered as an initial condition)
mean that the time dependent stress field can be written in terms
of integration of line integrals over the dislocation loop, 
consequently it is independent of the choice of slip planes.

\section{Examples for the stationary case}

\subsection{A screw dislocation in a half space}

Using the method in Eshelby and Stroh (1951) the displacement 
and stress fields in polar
coordinates due to a screw dislocation lying along the $z$ axis perpendicular
to the free surface of a half space, can be obtained as (the
half space exists for $z > 0$):
\begin{eqnarray}
U_{z}(\theta) & = & \frac{b}{2 \pi} \theta \nonumber \\
U_{\theta}(r,z) & = & \frac{b}{2 \pi} \int_{0}^{\infty} \frac{dk}{k}
e^{-k z} J_{1}(kr) \nonumber \\
\sigma_{z \theta}(r,z) & = & \frac{\mu b}{2 \pi r} -
\frac{\mu b}{2 \pi} \int_{0}^{\infty} dk e^{-k z}
J_{1}(kr) \nonumber \\
\sigma_{r \theta}(r,z) & = & -\frac{\mu b}{2 \pi}
\int_{0}^{\infty} dk e^{-k z} J_{2}(kr)
\label{sigh}
\end{eqnarray}
($J_{1}(z)$ the Bessel's function of order one).
Applying the general
formula of Eq. (\ref{fin}): 
\beq
\sigma_{z \eta}(\vec{R},z) = b \mu
\epsilon_{\alpha \eta} \int_{0}^{\infty}
dz' \sigma_{z \alpha}^{Gz}(z',z;-\vec{R})
-b \mu \int_{C} dR_{\beta}'
A_{\beta}^{z z \delta}(z'=0,z;\vec{R}'-\vec{R}) \: ,
\label{sze}
\eeq
with greek indices indicating coordinates $x$ or $y$ only, $\vec{R} \equiv
x \hat{i} + y \hat{j}$, $\mu$ the shear elastic constant, and $C$ is 
an arbitrary curve on the free surface.
The "vector potential" $A_{\beta}^{z z \delta}$ at the free surface is given
in Fourier space by:
\beq
A_{\beta}^{z z \delta}(z'=0,z;\vec{k})=
\frac{e^{-k z}}{k^{2}}k_{\beta}\epsilon_{\delta \eta} k_{\eta}
\: .
\eeq
If one calculates $\sigma_{z \theta}(r,z)$ using Eq. (\ref{sze}), one
reproduces the first and second terms in Eq. (\ref{sigh}) with the
equivalent terms in Eq. (\ref{sze}).
Gosling and Willis (1994) have given 
a different
line integral representation specifically determined for the geometry
of an isotropic half space. 
 
\subsection{A screw dislocation in a plate}

In Eshelby et al. (1951) the exact displacement and stress fields in polar
coordinates due to a screw dislocation along the $z$ axis perpendicular
to the free surfaces of a plate, were obtained as (the
plate exists for $-h < z < h$):
\begin{eqnarray}
U_{z}(\theta) & = & \frac{b}{2 \pi} \theta \nonumber \\
U_{\theta}(r,z) & = & -\frac{b}{2 \pi} \int_{0}^{\infty} \frac{dk}{k}
\frac{\sinh (kz)}{\cosh (kh)} J_{1}(kr) \nonumber \\
\sigma_{z \theta}(r,z) & = & \frac{\mu b}{2 \pi r} -
\frac{\mu b}{2 \pi} \int_{0}^{\infty} dk \frac{\cosh (kz)}{\cosh (kh)}
J_{1}(kr) \nonumber \\
\sigma_{r \theta}(r,z) & = & \frac{\mu b}{2 \pi}
\int_{0}^{\infty} dk \frac{\sinh (kz)}{\cosh (kz)} J_{2}(kr) \: .
\label{sigp}
\end{eqnarray}
The calculation of $\sigma_{z \theta}(r,z)$ 
using Eq. (\ref{fin}) is 
analogous to the half-space case.
One needs the "vector potential" $A_{\beta}^{z z \delta}$ evaluated at
the surface, which in Fourier space is:
\beq
A_{\beta}^{z z \delta}(\pm h,z;\vec{k})  = 
-\frac{k_{\beta}\epsilon_{\delta \nu} k_{\nu}}{2k^{2}}
\{ \pm \frac{\cosh(kz)}{\cosh(kh)}  +
\frac{\sinh(kz)}{\sinh(kh)} \} \: .
\eeq

\section*{Acknowledgments}

This work was supported in part by the Andes Foundation, Fondecyt
Grant 3950011 and a C\'atedra Presidencial en Ciencias. Useful
discussions with F. Lund are gratefully acknowledged.

\newpage

\section*{References}

\begin{itemize}
\item Boudet, J.F., Ciliberto S., and Steinberg J., J. Phys. II France, 6, 1493 
(1996).

\item Eshelby, J.D., and Stroh, A.N., Philos. Mag., Ser. 7, Vol.
42, 1401 (1951).

\item Gosling, T.J., and Willis, J.R., J. Mech. Phys. Solids,
Vol. 42, $n^{o}$ 8, 1199 (1994).

\item Lund, F., Phys. Rev. Lett., 76, 2742 (1996).

\item Mura, T., Philos. Mag. 8, 843 (1963).

\item Sharon, E., Gross, S.P, and Fineberg, J., 
Phys. Rev. Lett., 76, 2117 (1996).

\end{itemize}

\end{document}